\documentclass{article}

\usepackage{PRIMEarxiv}

\usepackage[utf8]{inputenc} 
\usepackage[T1]{fontenc}    
\usepackage{hyperref}       
\usepackage{url}            
\usepackage{booktabs}       
\usepackage{amsfonts}       
\usepackage{nicefrac}       
\usepackage{microtype}      
\usepackage{lipsum}
\usepackage{fancyhdr}       
\usepackage{graphicx}       
\graphicspath{{media/}}     

\usepackage{multirow}
\usepackage{array}
\newcolumntype{P}[1]{>{\centering\arraybackslash}p{#1}}
\newcolumntype{M}[1]{>{\centering\arraybackslash}m{#1}}
\usepackage{graphicx}
\usepackage{amsmath}
\usepackage{amssymb}
\usepackage{float}
\usepackage{tcolorbox}
\usepackage{xcolor}

\title{Quantifying the Value of Seismic Structural Health Monitoring for post-earthquake recovery of electric power system in terms of resilience enhancement
}

\author{
  Huangbin Liang\\
  Future Resilient Systems \\Singapore-ETH Centre \\
  Singapore\\
Corresponding email:\\ huangbin.liang@sec.ethz.ch\\
   \And
  Beatriz Moya \\
  ENSAM Institute of Technology \\
  Paris, France\\
  \\
  CNRS@CREATE LTD., CNRS \\
  Singapore\\
  \AND
   Francisco Chinesta\\
  ENSAM Institute of Technology \\
  Paris, France\\
    \\
  CNRS@CREATE LTD., CNRS \\
  Singapore\\
  \And
  Eleni Chatzi \\
  Department of Civil Environmental \\ and Geomatic Engineering \\
  ETH Zürich \\
  Zurich, Switzerland\\
}

\begin{document}
\maketitle

\begin{abstract}
Post-earthquake recovery of electric power networks (EPNs) is critical to community resilience. Traditional recovery processes often rely on prolonged and imprecise manual inspections for damage diagnosis, leading to suboptimal repair prioritization and extended service disruptions. Seismic Structural Health Monitoring (SSHM) offers the potential to expedite post-earthquake recovery by enabling more accurate and timely damage assessment. However, the deployment of SSHM comes with a cost and the quantifiable benefit of SSHM in terms of system-level resilience remains underexplored. This study develops an integrated probabilistic simulation framework to quantify the system-level value of SSHM in enhancing EPN resilience. The framework incorporates damage simulations based on EPN configuration, seismic hazard, fragility function, and damage-functionality mapping models, along with recovery simulations considering repair scheduling, resource constraints, transfer and repair durations. System functionality is evaluated via graph-based island detection and optimal power flow analysis under electrical constraints. Resilience is quantified using the Lack of Resilience (LoR) metric derived from the time-evolution functionality restoration curve. The effect of SSHM is incorporated by altering the quality of damage information used to create repair schedules. Specifically, different monitoring scenarios (e.g., no-SSHM baseline, partial SSHM, and full SSHM with various assessing accuracy levels) are modelled using confusion matrices that simulate misclassification of component damage states. The results demonstrate that improved damage awareness enabled by SSHM significantly accelerates recovery and reduces LoR by up to 21\% across various monitoring conditions. This study provides a quantitative foundation for evaluating the system-level resilience benefits of SSHM and guiding evidence-based sensor investment decisions for critical infrastructures.

\end{abstract}



\section{Introduction}
Electric power networks (EPNs) are the backbone of modern society, providing essential energy services to residential, industrial, commercial, and emergency sectors. However, they are increasingly exposed to natural hazards, earthquakes in particular, due to the distributed nature of their components across wide geographical areas that often intersect active fault zones \cite{fujisaki2014seismic}. Earthquake-induced EPN component failures can lead to widespread outages and cascading malfunctions across other interdependent lifeline systems, posing threats to public safety and economic continuity \cite{ouyang2014review,guo2017critical}. Consequently, enhancing the seismic resilience of EPNs, by improving their ability to absorb disruptions and rapidly recover functionality, has become a critical priority in the domain of disaster risk management and infrastructure planning \cite{espinoza2020risk,liang2023seismic,oboudi2024two}. However, post-earthquake recovery decisions are often hindered by limited situational awareness, resulting from delayed and inaccurate inspection-based assessment of component-level damage \cite{galloway2014lessons}. This challenge is particularly acute in large-scale power systems, where manual inspections are labor-intensive, time-consuming, and prone to errors in emergency settings \cite{kongar2017post}.

Over the past two decades, numerous frameworks have been developed to assess and improve the seismic resilience of EPNs \cite{oboudi2024two,babu2025comprehensive,xie2025resilience}, which integrate component-level fragility modeling, system-level functionality assessment, and post-event recovery simulation. Early studies primarily focused on evaluating physical vulnerabilities of individual components, such as power plants, transmission tower-lines \cite{miguel2021performance}, or substations \cite{liang2021system}, based on fragility curves and seismic hazard exposure. This component-level assessment approach was later extended to system-level frameworks, wherein the impact of localized damage on network-wide performance was evaluated via network connectivity methods or power flow models \cite{espinoza2020risk,xie2025resilience,cavalieri2014models}. These approaches typically integrate ground motion prediction equations (GMPEs) and Monte Carlo simulations (MCS) to generate damage scenarios and quantify the expected functionality loss. Building on damage simulation models, recent research has further shifted toward the dynamic process of post-earthquake recovery. Agent-based models and optimization-based scheduling algorithms have been proposed to simulate and optimize repair schedules under constraints such as limited crews, access delays, and interdependencies with other infrastructures \cite{oboudi2024two,xie2025resilience,sun2019agent,liu2023post,liang2022resilience}. Despite the progress in seismic resilience modeling, existing frameworks assume complete and accurate knowledge of component damage immediately after an earthquake \cite{liang2023seismic,oboudi2024two,babu2025comprehensive,xie2025resilience}. However, in real-world post-earthquake scenarios, such information is rarely available. Component-level damage must typically be inferred from visual inspections conducted by field personnel, which are time-consuming and often subject to human judgment[8]. These subjective assessments often deviate significantly from the actual damage conditions, potentially leading to suboptimal or even erroneous recovery decisions. For instance, if a severely damaged substation is mistakenly assessed as functional due to oversight or limited access, it may be excluded from early repair efforts, resulting in prolonged outages and greater system-wide functionality loss. Conversely, misidentifying intact components as damaged can lead to inefficient resource allocation. Thus, the epistemic uncertainty associated with post-earthquake inspections must be explicitly accounted for in resilience modeling to ensure realistic evaluation of recovery performance.

To address these limitations, Seismic Structural Health Monitoring (SSHM) has emerged as a promising solution for enhancing post-earthquake situational awareness \cite{limongelli2019seismic,karakostas2024seismic}. These systems typically employ a network of sensors—including accelerometers, strain gauges, and displacement transducers—integrated into critical infrastructure components. By continuously collecting measurements of structural responses during and after earthquakes, SSHM enables rapid post-event damage detection through signal processing techniques such as modal identification, frequency shifts, and pattern recognition \cite{karakostas2024seismic,quqa2021seismic,reuland2023comparative,figueiredo2022three}. Recent advances in data assimilation and digital twinning technologies have further enhanced SSHM by delivering virtual replicas of physical assets that may be updated dynamically as new sensing data arrives, supporting timely diagnostics and scenario-based decision-making \cite{malekloo2022machine,torzoni2024digital,liang2025harnessing}, and thus reducing the reliance on subjective and delayed field inspections. Several studies have demonstrated the feasibility and advantages of SSHM for post-earthquake applications across various engineering structures. For instance, Zhang et al. \cite{zhang2024post} proposed a framework for quantifying structural damage and evaluating damage state in RC frame structures leveraging SSHM data. Giordano et al. \cite{giordano2022value} further evaluated the role of SSHM in informing building evacuation decisions and concluded that timely structural information can reduce uncertainty and enhance occupant safety. Similarly, Ge et al. \cite{ge2021rapid} and Giordano \& Limongelli \cite{giordano2022value2} investigated sensing-based damage assessment and operational decisions for bridges. In the power system domain, Moya et al. \cite{moya50post} and Zhu et al. \cite{zhu2024post} explored the use of hybrid modelling based and pure data-driven anomaly detection approaches respectively for interconnected equipment in substations. But these studies remain largely confined to academic research and are decoupled from comprehensive post-disaster recovery contexts.

On the other hand, the adoption of SSHM is not without cost. The deployment and maintenance of dense sensing infrastructure require significant financial investment. This raises a fundamental question in infrastructure management: under which condition is it worthwhile to invest in SSHM to support post-earthquake recovery of EPNs? Value of Information (VoI) analysis has emerged as a formal framework to address this question \cite{straub2017value}. Rooted in Bayesian decision theory, VoI quantifies the expected gain from improved information provided by the SSHM when making decisions under uncertainty \cite{straub2017value}. Previous applications of VoI have included optimizing inspection and maintenance schedules under long-term environmental effects \cite{straub2014value,kamariotis2023framework,kamariotis2022value,zhang2022voi}, improving post-disaster assessments of building evacuation needs and decisions on bridge closures \cite{giordano2022value,giordano2022value2,giordano2023quantifying}, and evaluating the SHM system arrangement for structures \cite{zhang2022voi3,giordano2023value,long2020information}. However, these studies primarily address isolated structures and rely on relatively simple decision-making frameworks, typically involving binary choices such as whether to keep a bridge open or closed. Extending such insights to a quantitative evaluation of SSHM's value in supporting resilience-oriented, system-wide recovery planning for EPNs is far from straightforward. This complexity arises from the spatially distributed and functionally interdependent nature of EPN components \cite{wang2022systematic}, where localized errors in damage estimation can propagate through the network and significantly influence downstream repair sequencing decisions. A few notable exceptions have attempted to incorporate SHM data into probabilistic risk models or network management strategies \cite{liang2025harnessing,argyroudis2022digital,makhoul2024seismic}. However, these efforts are typically limited to conceptual illustrations and do not incorporate comprehensive system recovery modeling. Such modeling requires explicit consideration of sequential repair actions, component-specific repair durations, and constraints on available repair resources; elements that are critical for realistically capturing the dynamics of post-earthquake recovery. The Omission of these elements hampers realistic evaluation of how restoration sequencing impacts recovery efficiency and system-wide resilience, and as a result, the quantifiable benefit of SSHM in terms of system resilience remains underexplored. Specifically, no work has considered how imperfect, partial, or delayed monitoring affects recovery sequencing decisions and the resulting resilience outcomes. The integration of SSHM for post-earthquake recovery of EPNs in terms of resilience enhancement—especially with respect to evaluating trade-offs associated with sensing accuracy, coverage, and timeliness—remains an open and insufficiently addressed area.

To address these gaps, this study introduces a novel simulation-based framework to quantify the resilience-based value of SSHM information in supporting post-earthquake recovery of EPNs. The framework consists of four sequentially integrated modules: (1) a probabilistic damage simulation module; (2) a component-wise damage perception module that captures imperfect, partial, and delayed information; (3) a system-level recovery simulation module; and (4) a resilience-based VoI quantification module. This modular structure enables explicit evaluation of how SSHM influences repair scheduling, functionality recovery trajectories, and key resilience metrics, with uncertainties propagated throughout the modeling chain via Monte Carlo simulation.

A case study based on the IEEE 24-bus Reliability Test System demonstrates the application of the framework. System functionality is assessed through graph-based island detection and optimal power flow analysis under electrical constraints, while resilience is quantified using the Lack of Resilience (LoR) derived from the system’s time-dependent functionality restoration curve. By comparing scenarios with and without SSHM under identical hazard realizations, the resilience gains attributable to SSHM-enabled perception are quantified. Results show that SSHM significantly accelerates recovery and reduces LoR. Furthermore, sensitivity analyses investigate how variations in SSHM configurations—such as sensing accuracy and spatial coverage—affect recovery performance and resilience outcomes. These findings provide a quantitative foundation for guiding SSHM investment decisions and for supporting informed, uncertainty-aware recovery planning.

\section{Methodology}
\begin{figure}
    \centering
    \includegraphics[width=1\linewidth]
    { 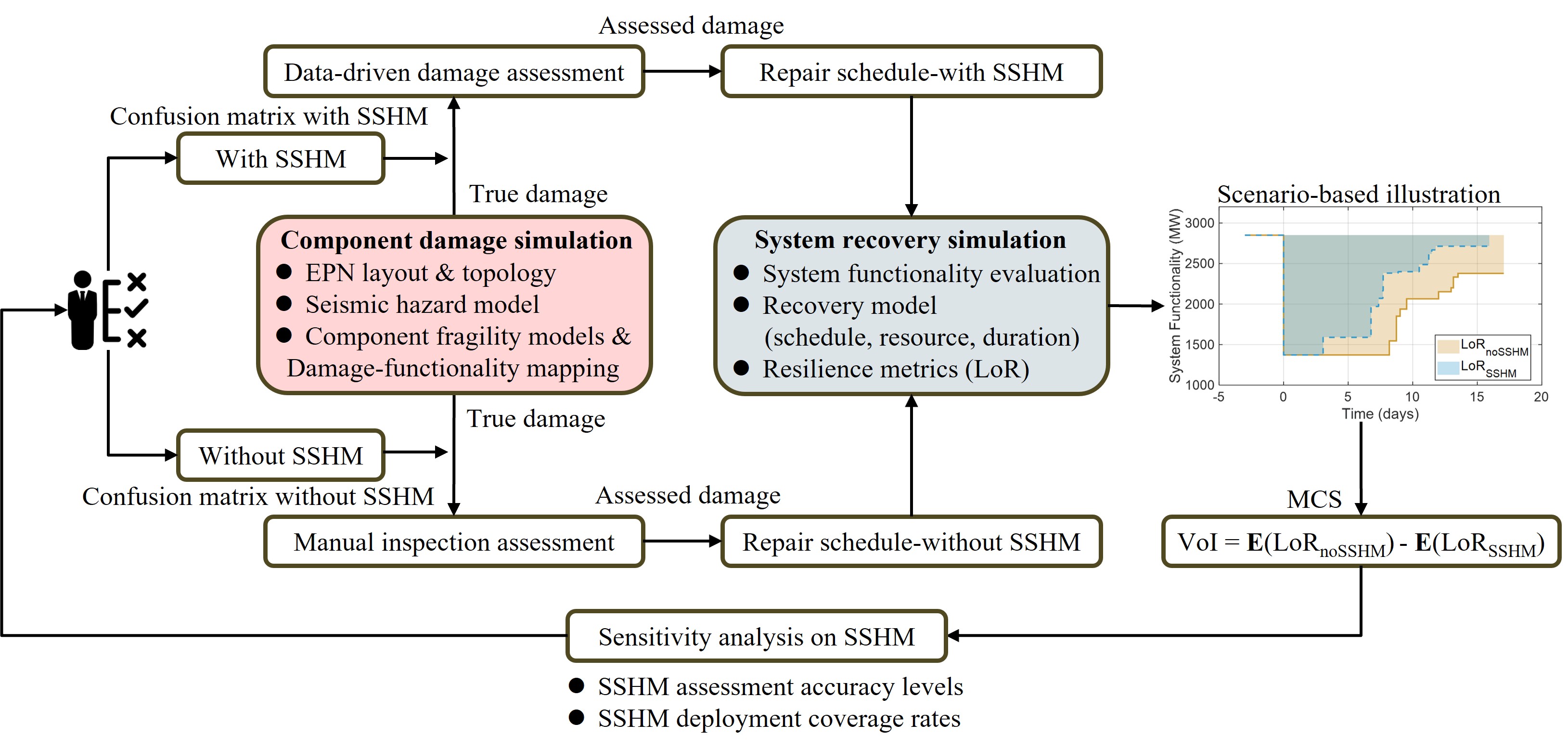}
    \caption{Framework for quantifying the value of SSHM information in supporting post-earthquake recovery of EPNs in terms of resilience enhancement. The framework integrates four modules: (1) component damage scenario generation using spatially correlated ground motions and fragility functions; (2) SSHM-based damage perception modeling with adjustable coverage, accuracy, and latency via confusion matrices; (3) system recovery simulation that schedules repairs based on perceived damage states and tracks system functionality over time; and (4) VoI quantification to assess resilience enhancement and uncertainties through Monte Carlo analysis.}
    \label{framework}
\end{figure}

This section outlines the proposed framework for quantifying the resilience-oriented value of SSHM information in supporting post-earthquake recovery of EPNs. As shown in Figure \ref{framework}, the framework comprises four tightly integrated modules: (1) a probabilistic damage simulation engine that generates earthquake-induced component damage scenarios using spatially correlated ground motion fields and fragility functions; (2) a component-wise damage perception module that simulates SSHM performance under varying coverage, accuracy, and latency assumptions, employing confusion matrices to model damage misclassification probabilities; and (3) a recovery simulation module that schedules repair actions based on the perceived (assessed) damage states and tracks the evolution of system functionality over time; and (4) a VoI quantification module that estimates the expected resilience enhancement and quantifies the associated uncertainties via Monte Carlo simulations. This integrated framework enables a comprehensive assessment of how SSHM influences repair scheduling, system functionality recovery trajectories, and resilience outcomes under realistic uncertainty. The following subsections detail the modeling assumptions and procedures of each module.

\subsection{Probabilistic damage simulation engine}
\begin{figure}
    \centering
    \includegraphics[width=1\linewidth]
    { 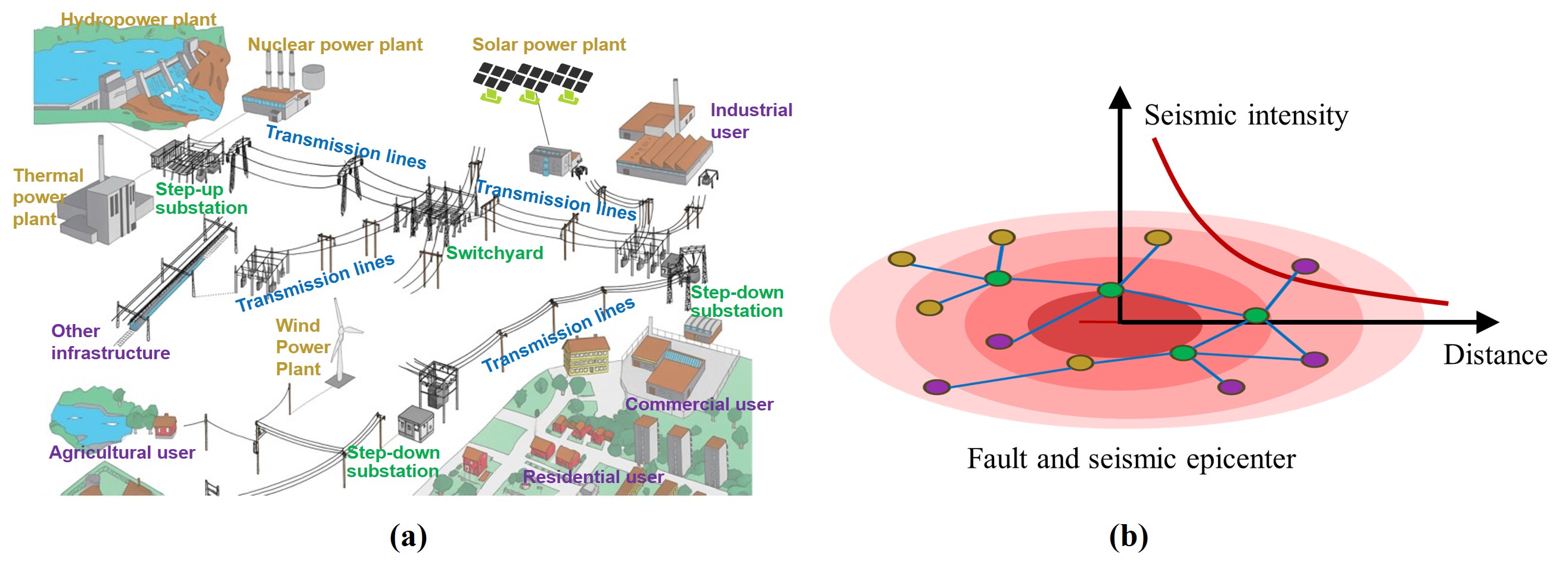}
    \caption{Schematic illustration of the earthquake impact on the EPN: (a) spatial layout of EPN components, including generation plants, transmission lines, substations, and end-users (adapted from \cite{xie2022research}); (b) conceptual representation of ground motion intensity attenuation from the earthquake epicenter, overlaid on the graph-based network topology to illustrate how spatially varying seismic demands interact with EPN components (nodes).}
    \label{EPN_ilustration}
\end{figure}

This module generates realistic component-level damage scenarios by integrating spatially correlated ground motion simulations with EPN system topology and component fragility functions. Figure \ref{EPN_ilustration}(a) provides an overview of a representative electric power network (EPN) distributed over a wide geographical area, including generation sources (hydropower, thermal, nuclear, solar, wind), electrical substations, transmission lines, and diverse user sectors (residential, commercial, industrial, agricultural). The illustration shows the spatially distributed and interconnected nature of EPNs, whose topology can be effectively represented as a graph $\mathcal{G}$(V, E) consisting of nodes (e.g., generators, substations, load units) and edges (transmission lines), enabling follow-up quantitative analysis of network connectivity and cascading effects resulting from component failures under seismic excitation.

The probabilistic damage simulation begins by modeling spatially correlated ground motion intensity fields. These are generated based on regional seismic source characteristics, fault geometry, and empirical ground motion prediction equations (GMPEs), which estimate seismic intensity measures such as peak ground acceleration (PGA) or peak ground velocity (PGV) at each component location:
\begin{equation}
    \ln(IM_i) = f(M, R_i, S_i) + \varepsilon_i \sigma
    \label{eq1}
\end{equation}
where $IM_i$ is the intensity measure at component $i$; $f$(·) is the median GMPE function dependent on earthquake magnitude M, source-to-site distance $R_i$, and local site conditions $S_i$; $\sigma$ is the standard deviation of logarithmic residuals, and $\epsilon_i$ is a spatially correlated standard normal variable capturing the intra-event variability across sites. To model the spatial correlation of residuals $\epsilon_i$, random fields are generated using empirically derived spatial correlation models (e.g., the model proposed by Jayaram and Baker \cite{jayaram2009correlation}), ensuring that intensity measures co-vary realistically between nearby components. 

The generated intensity fields are overlaid onto the EPN layout, as illustrated in Figure \ref{EPN_ilustration}(b), which conceptually shows the attenuation of ground motion intensity with increasing distance from the epicenter and assigns site-specific intensity measures to each EPN component. Component fragility functions are then used to probabilistically map these intensity measures to discrete damage states (e.g., none, slight, moderate, extensive, complete), by computing the conditional probability of exceeding damage thresholds given the assigned shaking intensity and specific component type. A typical fragility function for an EPN component can be represented in the lognormal form \cite{liang2025probabilistic,federal2012hazus,liu2021quantifying}:
\begin{equation}
    P(DS \geq ds_i | IM) = \Phi \left( \frac{\ln(IM) - \ln(\theta_i)}{\beta_i} \right)
    \label{eq2}
\end{equation}
where $P(DS \geq ds_i)$ is the probability of exceeding damage state $ds_i$; $\Phi$(·) denotes the standard normal cumulative distribution function; $\theta_i$ and $\beta_i$ are the median and logarithmic standard deviation of the fragility curve for damage state $i$.

\begin{figure}
    \centering
    \includegraphics[width=0.9\linewidth]
    { 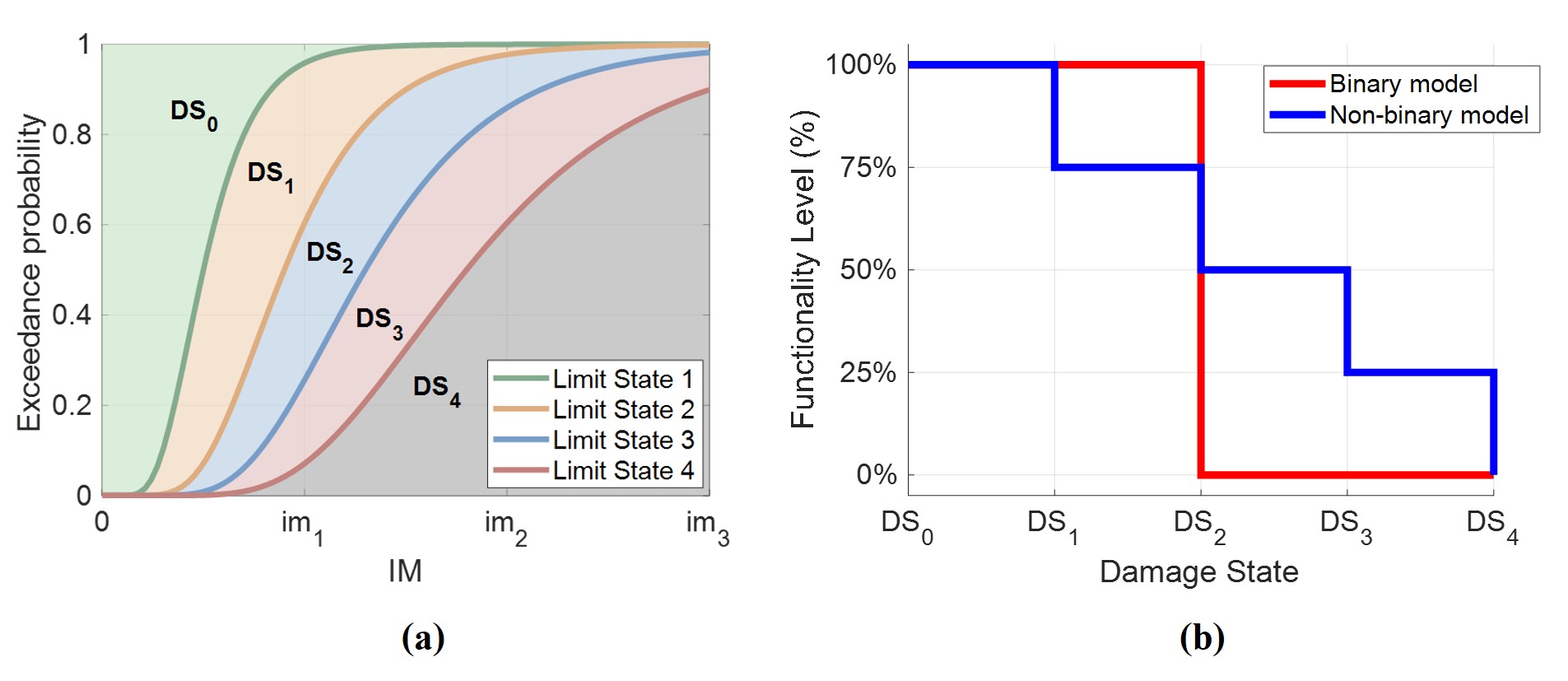}
    \caption{(a) Fragility curves illustrating the exceedance probabilities of different damage states (DS0–DS4) as functions of increasing intensity measures (IM), where vertical slices at given IM values define probability intervals for assigning component damage states; (b) mapping of discrete damage states to component functionality levels, showing binary models (red) for bus nodes and non-binary models (blue) for generation plants, load units, and substations with progressive capacity degradation.}
    \label{Fragility_curves}
\end{figure}

For each intensity field realization, Monte Carlo inverse sampling is performed: a random uniform number $u \sim \mathcal{U}(0,1)$ determines which damage state interval u falls into among the exceedance probabilities computed from the fragility functions, as illustrated in Figure \ref{Fragility_curves}(a). This probabilistic sampling process ensures that higher shaking intensities lead to greater probabilities of severe damage states. The resulting damage states are then translated into component functionality levels through predefined damage–functionality relationships, as depicted in Figure \ref{Fragility_curves}(b). Specifically, binary mappings are applied to bus nodes, which remain fully operational (100\%) in undamaged and slight damage states (i.e., DS$_0$ and DS$_1$) but become completely non-functional (0\%) beyond moderate damage (DS$_2$). Non-binary mappings are used for generation plants, load units, and substations, which experience progressive capacity loss under increasing damage severity (e.g., 75\%, 50\%, 25\% operability before complete failure). This differentiated damage–functionality mapping captures the distinct operational behaviors of various EPN components and enables a realistic assessment of cascading impacts in the system-wide functionality analysis.

\subsection{Component-wise damage perception module}
The component-level damage scenarios generated by the preceding probabilistic damage simulation module serve as the ground truth for subsequent assessment modeling. Accurate evaluation of these true damage states after the earthquake is essential for effective recovery planning. However, regardless of whether damage states are assessed through traditional inspection procedures or modern SSHM systems, damage perception inevitably involves uncertainty due to human interpretation errors, sensing limitations, or information delays. To account for these uncertainties, both SSHM-based and inspection-based damage assessments are modeled using a probabilistic confusion matrix framework that quantifies the likelihood of misclassifying damage states under different monitoring scenarios.

Let $D_i\in\{DS0, DS1, DS2, DS3, DS4\}$ denote the true damage state of component $i$, representing none, slight, moderate, extensive, and complete damage, respectively. The perceived damage state, denoted by $\hat{D}_i \in \{DS0, DS1, DS2, DS3, DS4\}$, may deviate from $D_i$ due to misclassifications inherent in the assessment process. We define a confusion matrix $\boldsymbol{C} \in \mathbb{R}^{5 \times 5}$ such that the element entry $\boldsymbol{C}_{m,n}$ represents the probability of perceiving damage level $DS_n$ given the true state $DS_m$:
\begin{equation}
    \boldsymbol{C}_{m,n} = P(\hat{D}_i = DS_n \mid D_i = DS_m), \quad \forall m,n \in \{0,1,2,3,4\}
    \label{eq3}
\end{equation}
where each row sums to one:
\begin{equation}
    \sum_{n=0}^{4} \boldsymbol{C}_{m,n} = 1, \quad \forall m \in \{0,1,2,3,4\}
    \label{eq4}
\end{equation}
Given a true damage state \(D_i = DS_m\) for component \(i\), the perceived state \(\hat{D}_i\) is modeled as a categorical random variable sampled from the probability vector given by the $m$-th row of $\boldsymbol{C}$ as follows:
\begin{equation}
    \hat{D}_i \sim \text{Categorical}(\boldsymbol{C}_{m,:})
    \label{5}
\end{equation}
This formulation captures the inherent misclassification risks in any post-earthquake assessment, whether SSHM-based or inspection-based, by probabilistically translating true damage states into potentially inaccurate assessed states that directly affect repair prioritization and system recovery strategies. To represent heterogeneous monitoring deployment scenarios, including full SSHM, partial SSHM coverage, and manual-only assessments, we introduce a monitoring indicator variable $\delta_i \in \{0, 1\}$, where  $\delta_i$  = 1 indicates component $i$ is assessed by SSHM (using confusion matrix $\boldsymbol{C}^{\text{SSHM}}$) and $\delta_i$ = 0 indicates assessment by manual inspection (using confusion matrix $\boldsymbol{C}^{\text{noSSHM}}$). Accordingly, the perceived damage state for each component is determined as: 
\begin{equation}
    \hat{D}_i = 
    \begin{cases} 
        \text{Categorical}(\boldsymbol{C}_{D_i}^{\text{SSHM}}), & \text{if } \delta = 1 \\
        \text{Categorical}(\boldsymbol{C}_{D_i}^{\text{noSSHM}}), & \text{if } \delta = 0
    \end{cases}
    \label{eq6}
\end{equation}
By systematically varying the elements of $\boldsymbol{C}^{\text{SSHM}}$ and the deployment configurations of $\delta_i$, this unified component-wise damage perception modeling framework enables sensitivity analyses on SSHM sensing reliability and deployment strategies, quantitatively evaluating their impacts on repair sequencing, functionality recovery trajectories, and overall system resilience.

\subsection{System-level recovery simulation module}
\begin{figure}
    \centering
    \includegraphics[width=1\linewidth]
    { 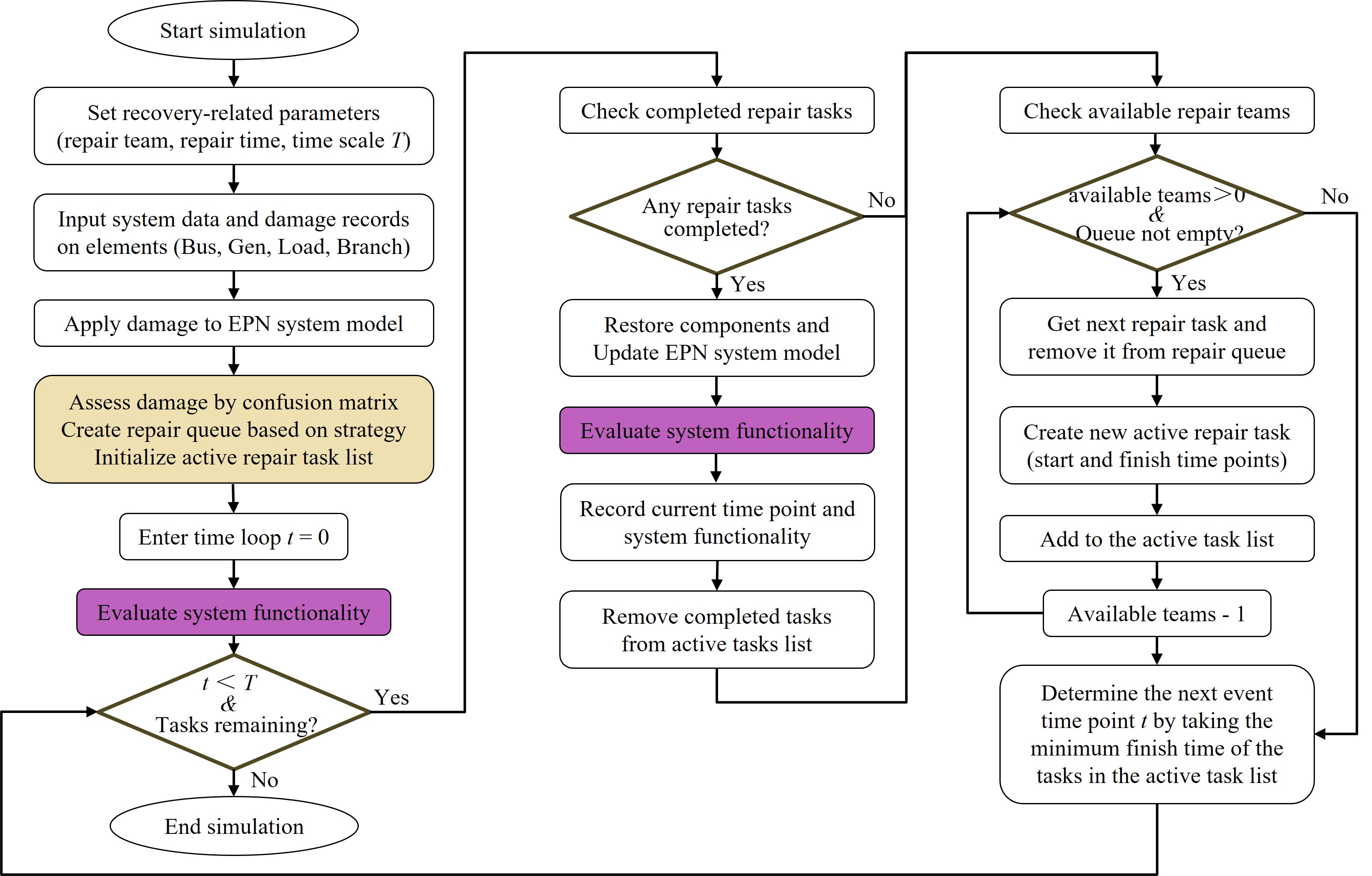}
    \caption{Event-driven flowchart of the post-earthquake EPN recovery simulation. The simulation begins with the creation of a prioritized repair queue and initialization of active repair tasks based on the perceived component damage states and repair strategy. In each simulation loop, completed repair tasks are checked and updated, new tasks are assigned to available crews considering transfer delays, and the simulation clock advances to the next scheduled event. After every repair completion, the system functionality is re-evaluated to capture network performance recovery gains. The simulation iterates until all repair tasks are completed or the maximum recovery time is reached. This dynamic workflow enables realistic modeling of time-dependent restoration trajectory, repair resource constraints, and the cascading impacts of component repairs on system-wide functionality.}
    \label{event-driven recovery flowchart}
\end{figure}

This module dynamically tracks the evolution of system functionality as damaged components are sequentially repaired. The simulation framework follows an event-driven architecture, as presented in Figure \ref{event-driven recovery flowchart}, which outlines the iterative process of checking completed repairs, assigning new repair tasks, updating system functionality, and advancing simulation time. 

The simulation begins by creating a prioritized repair queue based on the perceived damage states ${\hat{D}_i }$ and a selected repair strategy (e.g., damage severity-based, repair time-based, or criticality-based prioritization). An active repair task list is initialized, representing repair teams currently assigned to components. Each repair team is tracked for assignment and availability, with transition delays between repair sites explicitly modeled to reflect realistic mobilization. At each simulation timestep, the process involves: 1) check whether any active repair tasks completed by comparing their expected completion times to the current simulation time, and update the damage state of the corresponding component in the EPN system model accordingly; 2) evaluate whether new repair tasks can be dispatched given available teams and remaining tasks in the prioritized repair queue; 3) assign available teams to new tasks when they are free, setting their expected completion times based on component-specific repair durations; 4) update the next event time, which corresponds either to the earliest pending repair completion or the next crew becoming available; 5) advance the simulation clock to the next event time and repeat the process. This event-driven cycle continues iteratively until all repair tasks are completed or a maximum simulation time horizon is reached, with system functionality re-evaluated after every task completion event to capture the incremental restoration progress of the network. The system functionality evaluation, illustrated by the purple blocks in Figure \ref{event-driven recovery flowchart}, consists of a two-stage procedure.

(i) Island identification through graph-based topological analysis. As introduced in section 2.1, the network topology can be represented as a graph $\mathcal{G}$(V, E), where nodes correspond to buses, generation plants, or load units, and edges represent transmission lines or substations. Based on the current damage states, all failed nodes and disconnected edges are removed, and the updated network connectivity is encoded in an adjacency matrix $\mathbf{A} \in \{0,1\}^{n \times n}$: 
\begin{equation}
    A_{ij} = 
    \begin{cases} 
        1 & \text{if nodes } i \text{ and } j \text{ as well as edge } e_{ij} \text{ are intact} \\
        0 & \text{otherwise}
    \end{cases}
    \label{eq7}
\end{equation}
Electrically isolated subnetworks (islands) $\mathcal{G}_1$, $\mathcal{G}_2$, ..., $\mathcal{G}_N$ are identified using a breadth-first search (BFS) \cite{cormen2022introduction} traversal of $\mathbf{A}$. Each island is checked for operational viability: it must contain at least one functional generator and at least one load unit with nonzero demand; otherwise, it is considered inoperable and excluded from further analysis.

(ii) Power flow rerouting via DC optimal power flow (DCOPF). For each viable island, a DCOPF problem is formulated to determine how much load can be served under the current post-damage network configuration, accounting for degraded component capacities and operational constraints. The optimization problem for each island is given by:
\begin{equation}
\begin{aligned}
    & \min_{PG, \theta} \sum_{i \in G} C_i(PG_{i}) \\
    & \text{subject to} \\
    & \begin{cases}
    PG_i - PD_i = \sum_{j \in \mathcal{N}(i)} B_{ij} (\theta_i - \theta_j) & \text{(Nodal balance constraint)} \\
    |B_{ij}(\theta_i - \theta_j)| \leq \alpha_{ij} \cdot P_{\text{RATE},ij} & \text{(Line flow capacity constraint)} \\
    \alpha_i \cdot PG_i^{\text{min}} \leq PG_i \leq \alpha_i \cdot PG_i^{\text{max}} & \text{(Generator constraint)}
    \end{cases}
\end{aligned}
\label{eq8}
\end{equation}
where $PG_i$ is the generation output of generator $i$, with corresponding cost coefficient $C_i$; $PD_i$ and $\theta_i$ are the load demand and voltage angle at bus $i$, respectively; $ B_{ij}$ is the line admittance, and $P_{\text{RATE},ij}$ the transmission rate limit of line $(i, j)$; $\alpha_{ij}$ , $\alpha_j\in [0,1]$ denote the residual capacity ratio of lines and nodes, respectively, derived from their damage states and functionality mappings; $PG_i^{\text{min}}$ and $PG_i^{\text{max}}$ represent the minimum and maximum generation limits.

The total served load per island after solving the DCOPF is calculated as:
\begin{equation}
    F_{\mathcal{G}_k}^{\text{served}} = \sum_{j = \mathcal{G}_k} \text{PD}_j^{\text{served}}
    \label{eq9}
\end{equation}
where $\text{PD}_j^{\text{served}}$ is the actual load delivered to bus j after considering optimal flow dispatch and potential load shedding, in which the smallest non-zero load within the island is sequentially shed until power balance convergence is achieved or the retry limit is reached.

The system-level functionality at time $t$ is computed as the aggregate sum across all islands:
\begin{equation}
    F_t = \sum_{k=1}^{N} F_{\mathcal{G}_k}^{\text{served}}
    \label{eq10}
\end{equation}
which provides an instantaneous measure of the system performance during the recovery trajectory. By repeating this process at each repair event, a dynamic trajectory of system functionality over time can be generated, which explicitly accounts for the component-specific repair durations, the limited repair resources, and the cascading impacts of component repairs on system-wide functionality.

\subsection{Resilience-based VoI quantification module}
To evaluate whether investments in SSHM systems are justified, this final module quantifies the value of SSHM information by comparing the expected improvements in resilience metrics with and without SSHM-enabled damage perception under identical hazard scenarios. Specifically, resilience is measured using the widely adopted indicator, lack of resilience (LoR), quantified as the area between the pre-event functionality level and the evolving recovery trajectory over time:
\begin{equation}
    LoR = \int_{0}^{T} [F_0 - F(t)] \, dt
    \label{eq11}
\end{equation}
where $F_0$ denotes the initial (undamaged) system functionality, and $F(t)$ is the system functionality at time $t$. 

The VoI analysis is computed by comparing two expected resilience outcomes across a suite of Monte Carlo earthquake scenarios, each consisting of spatially correlated ground motion fields and corresponding true damage realizations. For every sampled scenario, the system recovery simulation is run twice: i) the prior recovery trajectory without SSHM, relying solely on traditional inspection procedures that have their own errors and uncertainties, and ii) the posterior recovery trajectory with SSHM-enabled perception, using confusion matrices with specified SSHM sensing accuracy and coverage levels. The expected value of information is then estimated as the mean improvement across the Monte Carlo samples:
\begin{equation}
  VoI = E \left[ LoR^{\text{noSSHM}} \right] - E \left[ LoR^{\text{SSHM}} \right]
  \label{eq12}
\end{equation}
where $E$[·] is the expectation operator, and positive VoI values indicate improvements attributable to SSHM deployment. The uncertainty associated with expected benefits is quantified using the standard deviation of the sample distributions of $LoR$, capturing variability introduced by stochastic ground motion fields and seismic fragility uncertainties. 

This integrated framework bridges the gap between purely technical performance metrics and actionable investment decisions in disaster resilience enhancement planning. Moreover, systematic VoI sensitivity analyses can be further performed by varying SSHM sensing accuracy (via confusion matrix elements) and coverage rates (proportion of instrumented components), to investigate how incremental investments in SSHM translate into resilience benefits, providing decision-makers with quantitative evidence on the marginal value of expanding SSHM deployments. These insights will be illustrated in the subsequent case study. 

\section{Case Study}
\subsection{Power network configuration}
\begin{figure}
    \centering
    \includegraphics[width=0.95\linewidth]
    { 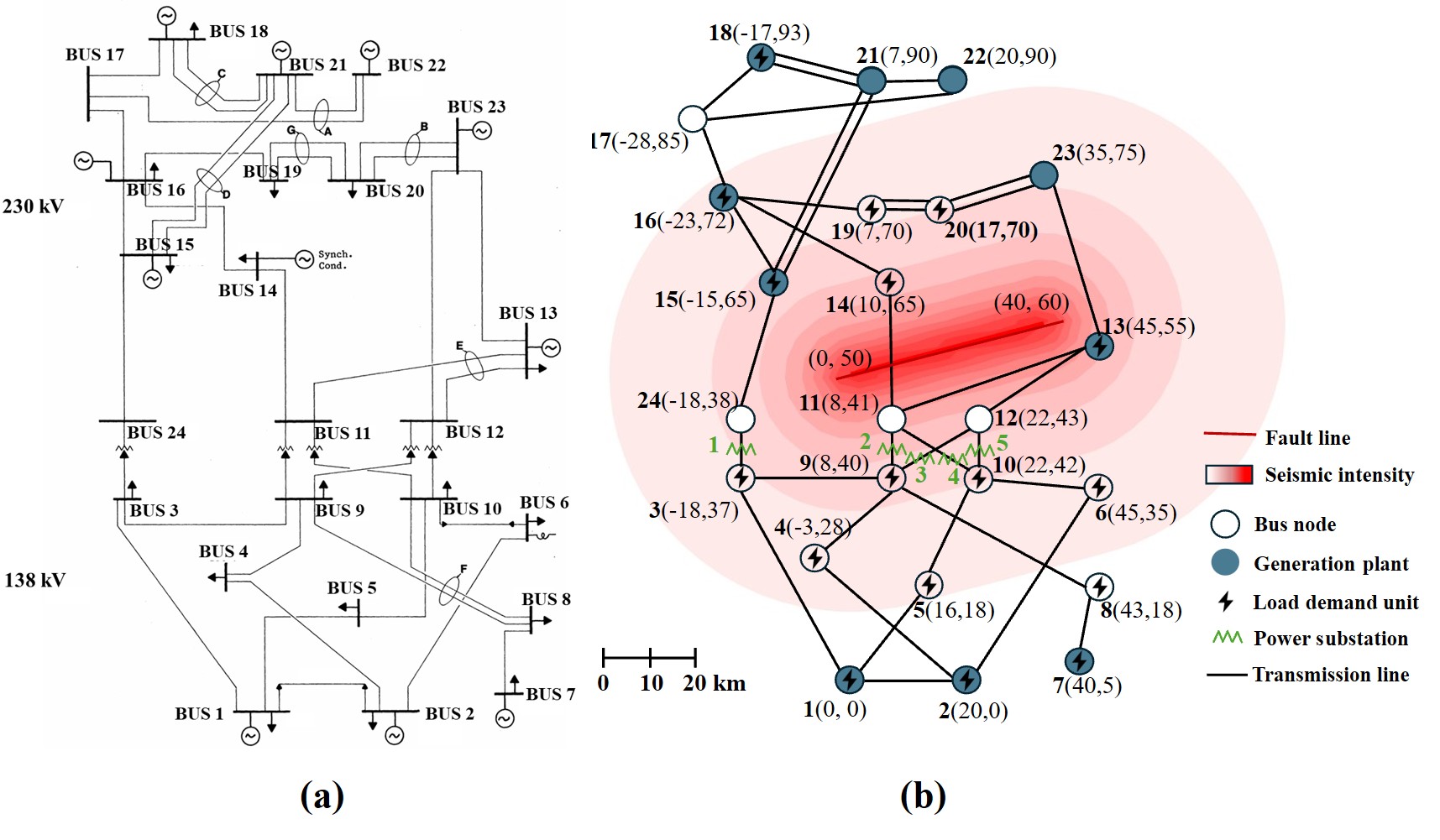}
    \caption{Illustrated IEEE 24-bus RTS (a) one-line diagram \cite{subcommittee1979ieee}; (b) abstracted network configuration diagram subjected to an earthquake scenario with a strike-slip fault line (dashed red) defined by two endpoints at coordinates (0, 50) and (40, 60) \cite{liang2025multi}. The red shaded contours represent the seismic intensity field, with darker zones indicating higher seismic intensity. The geographic coordinates of bus nodes (hollow circle) are indicated in parentheses. Generation plants, load units, and substations are denoted by blue circles, lightning symbols, and green polylines, respectively.}
    \label{IEEE 24-bus}
\end{figure}

To demonstrate the applicability and efficacy of the proposed framework, a comprehensive case study is conducted on the IEEE 24-bus Reliability Test System(RTS) \cite{subcommittee1979ieee}, which provides a standardized benchmark for resilience assessment of EPNs. As shown in Figure \ref{IEEE 24-bus}(a), the system spans two voltage levels (138 kV and 230 kV) with 24 buses connected by 38 transmission lines. The buses are categorized into generation-only, load-only, and mixed-function types, each characterized by operational parameters including voltage angles, generation capacities, and load demands, all obtained from the MATPOWER case data \cite{zimmerman2010matpower}. The single-line diagram can be abstracted into an undirected network graph of nodes and edges representing system components. Figure \ref{IEEE 24-bus}(b) illustrates this abstraction, depicting 10 generation plants (blue circles), 17 load centers (lightning symbols), and 5 substations (green polylines), all interconnected via bus nodes that act as key junctions for power injection, isolation, and routing. The baseline system functionality before an earthquake is 2850 MW. To simulate spatially distributed seismic impact, each bus node is assigned a 2D coordinate, as indicated in the parentheses in Figure \ref{IEEE 24-bus}(b).

\subsection{Seismic hazard modelling}
For illustrative purposes and without loss of generality, we define a hypothetical strike-slip fault crossing the central and northeast regions of the system, with endpoints at coordinates (0, 50) and (40, 60). The system is assumed to be situated in southwestern China with a representative hard soil site condition characterized by $VS_{30}$=760 m/s, and a synthetic earthquake scenario with a moment magnitude of 8.0 is adopted. In this study, peak ground acceleration (PGA) is selected as the seismic intensity measure, and ground motions at each component location are simulated by combining the BSSA14 GMPE attenuation model \cite{boore2014nga} and the spatial correlation model proposed by Jayaram and Baker \cite{jayaram2009correlation}, with a correlation length of 40 km to reflect realistic spatial variability. Accordingly, spatially correlated PGA fields can be generated, as shown in Figure \ref{IEEE 24-bus}(b) by the red shaded contours.

\subsection{Component fragility parameters}
The seismic fragility of key EPN components, including buses, generation plants, load units, and substations, is characterized using Equation \ref{eq2} conditioned on PGA. The corresponding median PGA thresholds ($\theta$) and logarithmic standard deviation ($\beta$) are sourced from the FEMA Hazus Earthquake Model \cite{federal2012hazus}, as summarized in Table \ref{tab:1}. Notably, Hazus does not provide fragility parameters for transmission towers or overhead lines, reflecting the common assumption in existing research that these structures possess inherent robustness against ground shaking due to their design for extreme environmental loads (e.g., wind, ice, cascading tower failures). Accordingly, this study assumes that transmission towers remain functional during earthquakes, supported by experimental studies \cite{liang2020shaking} and field evidence \cite{salman2017multihazard} indicating that damage to such structures typically arises from secondary hazards such as landslides or liquefaction, for which empirical data are limited.

\begin{table}[h]
    \centering
    \caption{Seismic fragility parameters for different EPN components}
    \begin{tabular}{lcccccccc}
        \toprule
        \textbf{Damage} & \multicolumn{2}{c}{\textbf{Bus node}} & \multicolumn{2}{c}{\textbf{Generation plant}} & \multicolumn{2}{c}{\textbf{Load unit}} & \multicolumn{2}{c}{\textbf{Substation}} \\
        \textbf{state} & $\theta(g)$ & $\beta$ & $\theta(g)$ & $\beta$ & $\theta(g)$ & $\beta$ & $\theta(g)$ & $\beta$ \\
        \midrule
        Slight & 0.13 & 0.65 & 0.10 & 0.60 & 0.24 & 0.25 & 0.10 & 0.60 \\
        Moderate & 0.26 & 0.50 & 0.22 & 0.55 & 0.32 & 0.23 & 0.20 & 0.50 \\
        Extensive & 0.34 & 0.40 & 0.49 & 0.50 & 0.58 & 0.15 & 0.30 & 0.40 \\
        Complete & 0.74 & 0.40 & 0.79 & 0.50 & 0.89 & 0.15 & 0.50 & 0.40 \\
        \bottomrule
    \end{tabular}
    \label{tab:1}
\end{table}

\subsection{Damage perception modelling}
To realistically capture the uncertainty related to damage perception during post-earthquake assessment, a unified tridiagonal confusion matrix is adopted for both inspection-based and SSHM-based assessments:
\begin{equation}
    \boldsymbol{C} = \begin{bmatrix}
    \frac{1+a}{2} & \frac{1-a}{2} & 0 & 0 & 0 \\
    \frac{1-a}{2} & a & \frac{1-a}{2} & 0 & 0 \\
    0 & \frac{1-a}{2} & a & \frac{1-a}{2} & 0 \\
    0 & 0 & \frac{1-a}{2} & a & \frac{1-a}{2} \\
    0 & 0 & 0 & \frac{1-a}{2} & \frac{1+a}{2}
    \end{bmatrix}
    \label{eq13}
\end{equation}
where the parameter $a \in$ [0,1] defines the sensing accuracy. The diagonal elements represent the probability of correctly identifying the true state, while off-diagonal elements in adjacent cells model the probability of under- or overestimating damage by one level. This narrow-band assumption is justified by the practical observation that fully intact and completely failed components are usually easier to distinguish, whereas moderate damage states often exhibit ambiguous features. As a result, misclassifications are more likely to occur between adjacent levels rather than across multiple damage states. For the case of inspection-based assessment, we set $a$ = 0.70 to represent typical field reconnaissance accuracy and apply a 2-day information delay before perceived damage data become available for repair scheduling to reflect inspection logistics. For SSHM-based assessments, we use $a$ = 0.90 to capture their higher precision, with immediate assessment data availability (0-day delay) representing real-time damage perception.

\subsection{Recovery-related parameters}
Three repair teams are assumed to be available for dispatch, each capable of repairing one damaged EPN component at a time. The mobilization delay between successive repairs is modelled as a fixed transfer time of 0.25 days, representing crew relocation across sites; while more sophisticated travel time models can be incorporated if detailed road network data are available.

Component repair durations are sampled from normal distributions parameterized according to damage severity and component type. Median repair times ($\mu$) and standard deviations ($\sigma$) are collected from the FEMA Hazus Earthquake Model \cite{federal2012hazus}, as detailed in Table \ref{tab:2}. To ensure physical plausibility, a lower bound of 0.2 days is imposed on sampled repair times, preventing negative or unrealistically short durations that could otherwise arise from the probabilistic modeling.

Repair prioritization follows a two-tiered strategy consistent with practical emergency restoration guidelines: first, components are ordered by type priority (bus nodes > generation plants > load units > substations), and second, within each type, repairs proceed by descending capacity (e.g., generation rating PG or load magnitude PD), enabling efficient restoration of critical elements and power supply paths to maximize early gains in system functionality.

\begin{table}[h]
    \centering
    \caption{Repair duration parameters for different EPN components (units: day)}
    \begin{tabular}{lcccccccc}
        \toprule
        \textbf{Damage} & \multicolumn{2}{c}{\textbf{Bus node}} & \multicolumn{2}{c}{\textbf{Generation plant}} & \multicolumn{2}{c}{\textbf{Load unit}} & \multicolumn{2}{c}{\textbf{Substation}} \\
        \textbf{state} & $\mu$ & $\sigma$ & $\mu$ & $\sigma$ & $\mu$ & $\sigma$ & $\mu$ & $\sigma$ \\
        \midrule
        Slight & 0.8 & 0.4 & 0.5 & 0.1 & 0.3 & 0.2 & 1.0 & 0.6 \\
        Moderate & 2.5 & 1.0 & 3.6 & 3.6 & 1.0 & 0.5 & 3.0 & 1.5 \\
        Extensive & 5.5 & 2.0 & 22.0 & 21.0 & 3.0 & 1.5 & 7.0 & 3.5 \\
        Complete & 7.0 & 3.0 & 65.0 & 30.0 & 7.0 & 3.0 & 30.0 & 15.0 \\
        \bottomrule
    \end{tabular}
    \label{tab:2}
\end{table}

\section{Results and Discussions}
\subsection{Seismic hazard modelling}
\begin{figure}
    \centering
    \includegraphics[width=0.8\linewidth]
    { 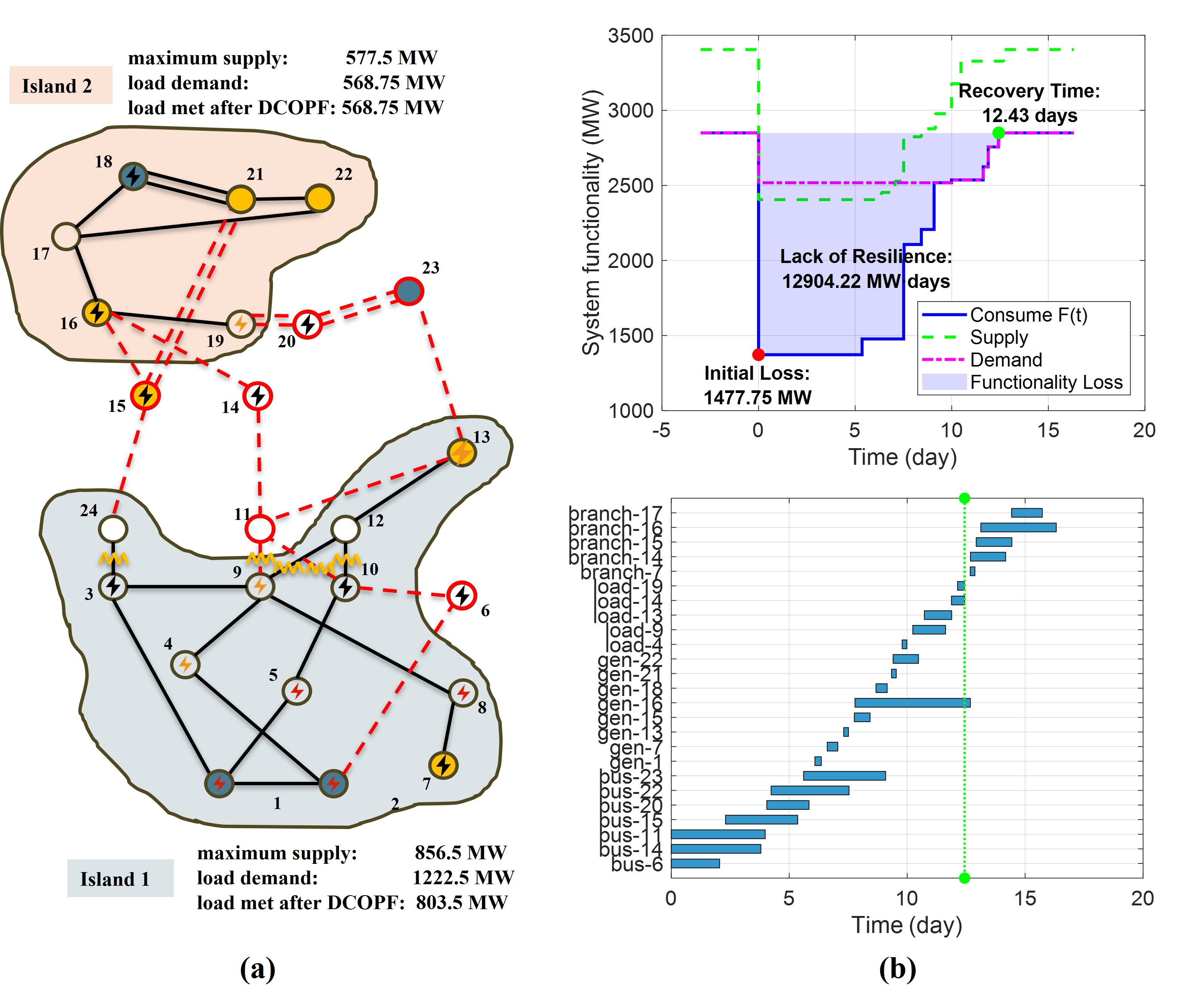}
    \caption{(a) Sampled damage state realization of EPN components after an earthquake. Components in red are completely damaged and must be isolated from the network, while orange indicates intermediate damage with partial functionality. The resulting configuration shows two identified electrical islands along with the DCOPF distribution results for this scenario; (b) recovery simulation results under perfect damage information: (Top) time-dependent system functionality curve; (Bottom) Gantt chart of component repairs.}
    \label{Sampled damage state}
\end{figure}

To establish a clear baseline understanding of how component-level damage perception errors affect recovery dynamics, we first evaluate one damage scenario paired with a single recovery simulation assuming perfect damage information. The illustrative sample realization of fragility-based damage states is visualized in Figure \ref{Sampled damage state}(a), where red components represent those completely damaged and must be isolated from the network, while orange components indicate intermediate damage states retaining partial functionality determined by the predefined damage–functionality mappings shown in Figure \ref{Fragility_curves}(b). After updating the network topology by removing failed nodes and edges, two active islands are identified through the graph-based BFS algorithm. Each island is then analyzed independently with DCOPF-based dispatch, implementing iterative load shedding until power balance is achieved. The results are also provided in Figure \ref{Sampled damage state}(a). As shown, Island 1 (buses 1–13, 24) meets 803.5 MW of 1222.5 MW demand, reflecting partial shedding due to electrical constraints, while Island 2 (buses 16–22) fully serves 568.75 MW of 568.75 MW demand. Consequently, the initial total system functionality sums to 1372.25 MW by Equation (\ref{eq10}).

Perfect damage information assumption is implemented by setting $a$=1 in the confusion matrix in Equation (\ref{eq13}), ensuring all components are classified without misclassification. Following the description in section 2.3, corresponding recovery simulation results are obtained and shown in Figure \ref{Sampled damage state}(b). As indicated in the bottom panel, there are a total of 25 components that experienced varying degrees of damage in the sampled scenario. The Gantt chart clearly shows the timing and sequence of these repairs based on the perfect damage information and the adopted recovery strategy, with each bar indicating the repair duration of a damaged component and the green vertical line marking the point of full system recovery. Notably, due to the complexity and redundancy of the EPN, and the multi-level functionality of individual components, full recovery to initial system functionality does not require every damaged component to be completely repaired. The top panel depicts the time-dependent performance trajectories, where the blue solid line indicates the evolution of the predefined system functionality $F(t)$, the green dashed line represents total supply capacity, and the magenta dashed-dot line shows total demand level. Despite the total generation capacity being sufficient to meet the aggregate load demand, the damaged network topology and power flow constraints limit actual load serviceability, resulting in substantial unmet demand during the early recovery period. The system initially experiences a functionality loss of 1477.75 MW, followed by gradual recovery, and reaching full restoration at approximately 12.43 days. The shaded area denotes the resilience metric LoR calculated as 12904 MW·day.

\begin{figure}
    \centering
    \includegraphics[width=1\linewidth]
    { 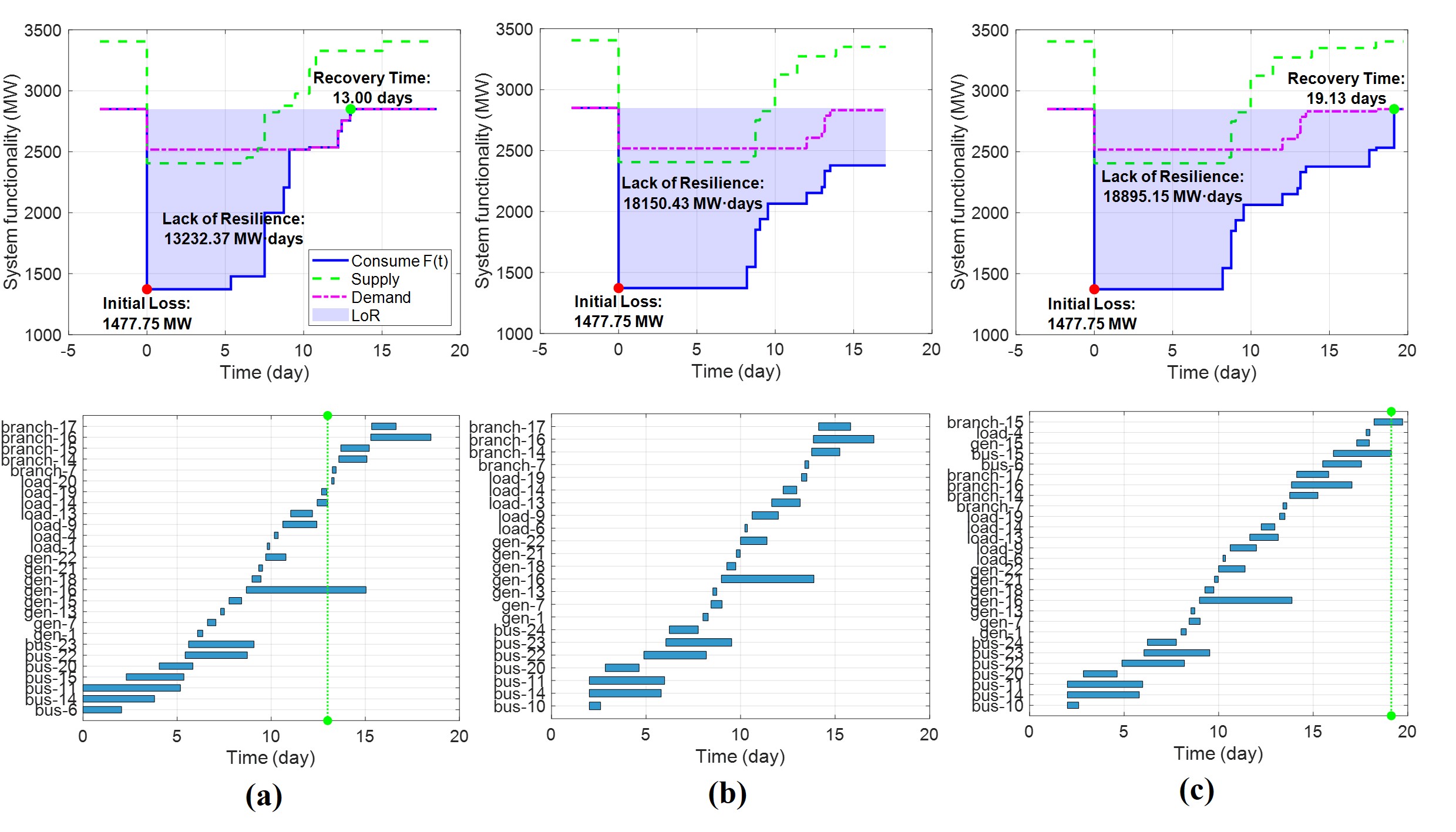}
    \caption{Sequencing decisions, recovery simulations, and resilience evaluation results under (a) SSHM-assisted damage perception scenario; (b) inspection-based damage perception scenario; (c) adjusted inspection-based scenario assuming missed damaged components are eventually detected and repaired after completing the initial repair list, with additional repair time penalties applied.}
    \label{Sequencing decisions}
\end{figure}

The impact of damage perception errors arising from inspection-based or SSHM-based assessments is analyzed using the perception modeling framework described in Sections 2.2 and 3.4. The resulting sequencing decisions, functionality trajectories, and resilience metrics under various damage conditions are presented in Figure \ref{Sequencing decisions}. Figure \ref{Sequencing decisions}(a) shows results for the SSHM-assisted case, where repair sequencing decisions are informed by relatively accurate and timely sensing, yielding a recovery time of 13.00 days and a LoR of 13232 MW·day, closely matching those under perfect damage information (12.43 days and 12904 MW·day), with only minor degradation due to the erroneous identification of undamaged components (specifically load 1 and load 20) as damaged.

In contrast, Figure \ref{Sequencing decisions}(b) presents the inspection-based scenario, where the recovery trajectory fails to restore system functionality to its original level even after completing all scheduled repair tasks. This discrepancy arises from the limited accuracy in inspection-based damage assessments, which not only mistakenly identify undamaged components (i.e., bus 24 and load 6) as damaged, but more critically, overlook several severely damaged components (i.e., bus 6 and 15, gen 15, load 4, and branch 15) by comparing with the repair schedule under perfect damage information presented in Figure \ref{Sampled damage state}(b). Such omissions lead to unrepaired faults persisting in the network, preventing the full recovery. To reflect more realistic operational practices, we further assume that overlooked damaged components are eventually discovered and repaired, albeit only after completing the initially scheduled repair tasks. Additionally, their repair durations are increased by 30\% to account for delays associated with double detection. As shown in Figure \ref{Sequencing decisions}(c), this adjusted scenario results in a final recovery time of 19.13 days and a substantially higher LoR of 18895 MW·day compared to the SSHM-assisted case, representing a resilience improvement of 5663 MW·day attributable to the timely and accurate sensing provided by SSHM, and underscoring its critical role in accelerating post-earthquake recovery.

\subsection{Single damage scenario with multiple recovery simulations}
\begin{figure}
    \centering
    \includegraphics[width=1\linewidth]
    { 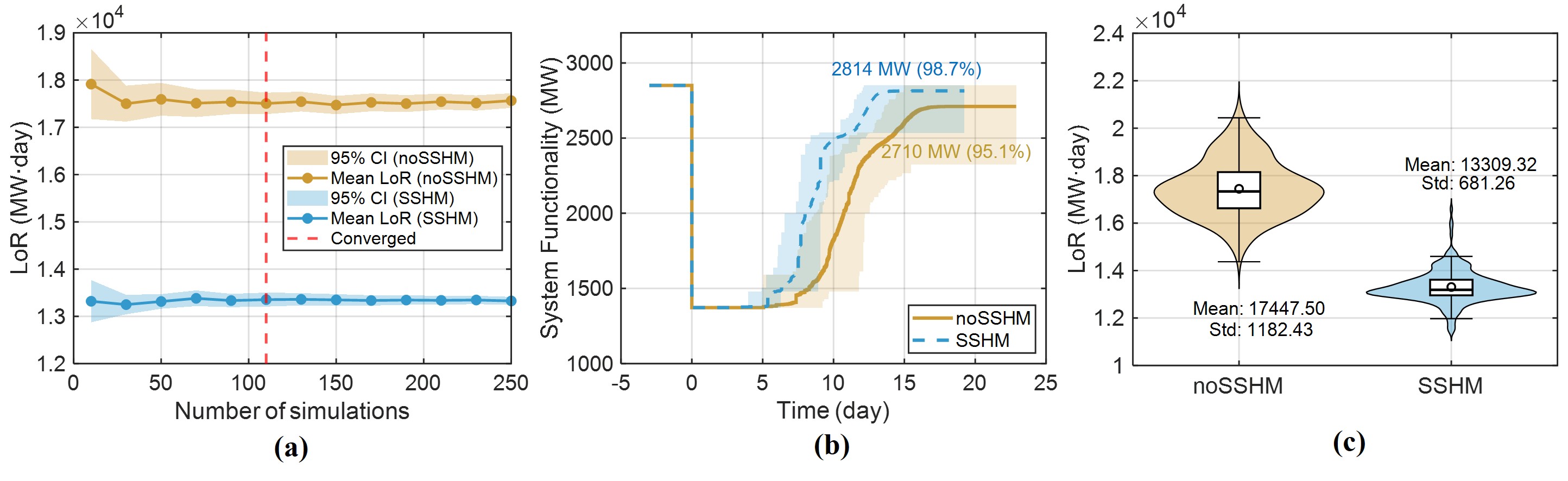}
    \caption{(a) Convergence behavior of the sample mean LoRs and their 95\% confidence intervals with the increasing MC simulations; (b) comparison of system functionality trajectories between inspection-based (noSSHM) and SSHM-assisted damage perception across the MC simulations; (c) comparison of the LoR distributions and statistics across the MC simulations.}
    \label{Convergence behavior}
\end{figure}

To evaluate the variability in recovery outcomes stemming solely from uncertainties in damage perception, multiple recovery simulations were performed based on the confusion matrix described in Section 3.4 under the same sampled damage scenario as Section 4.1. A total of 250 Monte Carlo (MC) realizations were conducted for both the SSHM-assisted and non-SSHM scenarios. Figure \ref{Convergence behavior}(a) shows the evolution of the mean LoR, and its 95\% confidence interval (CI) as the number of simulations increases. Convergence was achieved when the running average changed by less than 1\% over the last 20 simulations and the 95\% CI narrowed to within 2\% of the mean, which occurred after 110 simulations in both scenarios, confirming that the sample size is sufficient to provide statistically reliable estimates. 

Figure \ref{Convergence behavior}(b) displays the time evolution of mean system functionality trajectories for the inspection-based scenario (noSSHM, solid orange line) and SSHM-assisted scenario (dashed blue line) across 250 MC simulations, with shaded areas representing their respective 95\% CIs. As shown, under a single damage scenario, uncertainties in damage perception alone can significantly affect recovery trajectories. By comparison, the SSHM-supported case achieves faster and more consistent recovery with narrower uncertainty bands, whereas inspection-based assessment leads to slower and more variable restoration due to higher perception errors. Specifically, the SSHM scenario restores the system to a final functionality level of 2814 MW on average (98.7\% of pre-event capacity), while the inspection-based assessment scenario reaches only 2710 MW (95.1\%).

The distributions of the LoR values for scenarios with and without SSHM support are compared in Figure \ref{Convergence behavior}(c). Probability densities of LoR outcomes across the simulations are shown by violin plots, with internal boxplots indicating the median, interquartile range, and extreme values. A pronounced shift toward lower LoR values with reduced dispersion is evident in the SSHM-assisted case, highlighting substantial resilience benefits from accurate and timely damage perception provided by the SSHM. Quantitatively, a mean LoR of 13309 MW·day is derived with SSHM, compared to 17447 MW·day without SSHM, leading to a VoI of 4138 MW·day (23.7\%) in expected resilience improvement by Equation (\ref{eq12}). Moreover, the standard deviation of LoR is also reduced significantly from 1182 to 681 MW·day with an approximate 42.4\% reduction after employing SSHM, which underscores the enhanced consistency in recovery performance enabled by SSHM-assisted damage perception. 

\subsection{Multiple damage scenario with multiple recovery simulations}
\begin{figure}
    \centering
    \includegraphics[width=0.75\linewidth]
    { 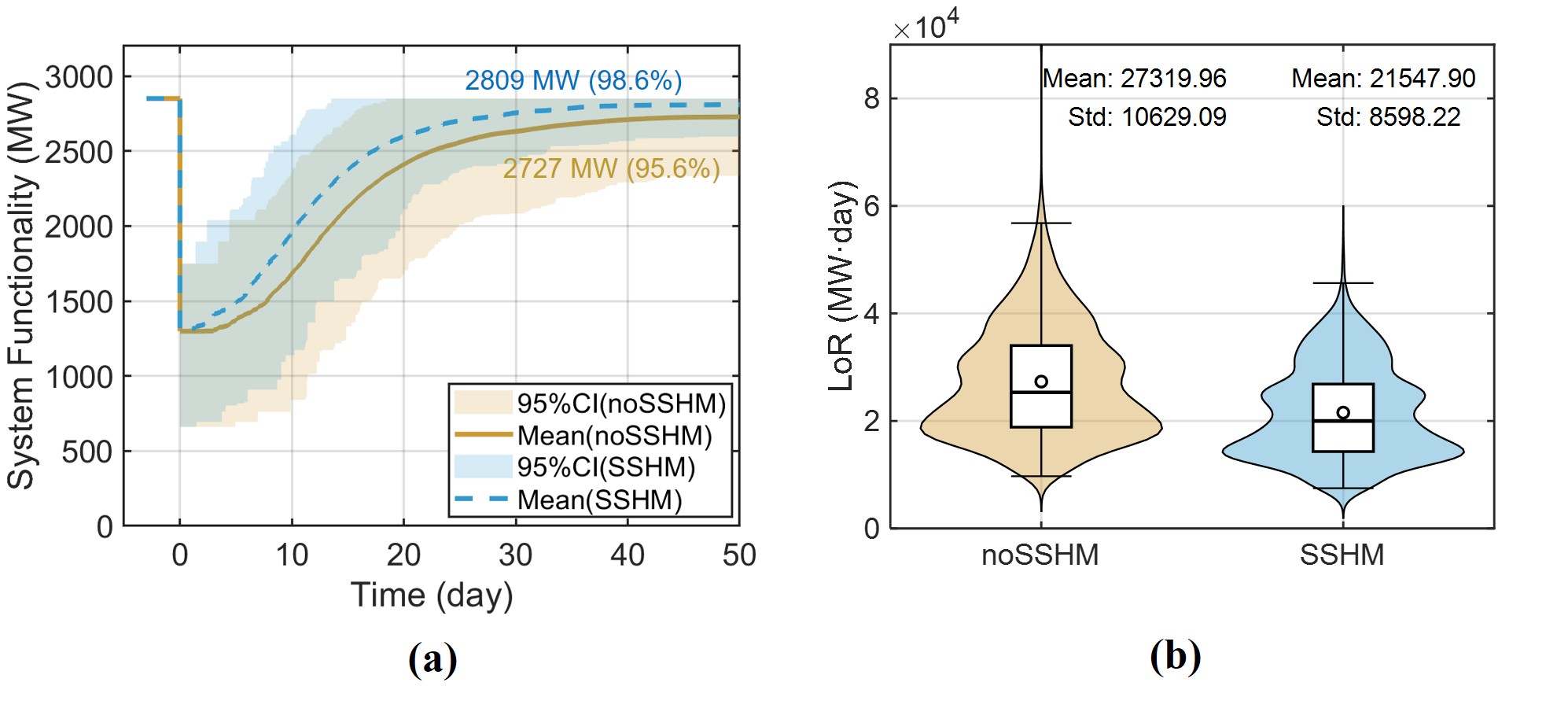}
    \caption{(a) Time evolution of mean system functionality with 95\% confidence intervals across the 10000 MC simulations; (b) violin plots of LoR distributions highlighting reduced mean and variability with SSHM-assisted damage perception compared to inspection-based assessment.}
    \label{10000 MC simulations}
\end{figure}

To capture the combined effects of hazard-induced damage variability and damage perception uncertainties, a set of 100 independent damage scenarios was generated using stochastic ground motion fields and component-specific fragility functions introduced in sections 3.2 and 3.3. For each scenario, 100 recovery simulations were performed under both inspection-based and SSHM-assisted perception assumptions, resulting in a total of 10000 recovery realizations per perception method. The results are summarized in Figure \ref{10000 MC simulations}. As shown in Figure \ref{10000 MC simulations}(a), further accounting for hazard-induced damage variability introduced a larger uncertainty to the recovery simulations compared with those shown in Figure \ref{Convergence behavior}(b), and SSHM-supported assessments consistently enable faster recovery with reduced uncertainty, indicating improved consistency in performance across diverse earthquake impacts. In addition, the final system functionality stabilizes at an average of 98.6\% of the pre-event functionality level, which is notably higher than the inspection-based assessment case (95.6\%). This 3.0\% improvement in ultimate functionality is attributable to more accurate damage perception, underscoring the critical role of SSHM in facilitating complete system restoration. 

Figure \ref{10000 MC simulations}(b) displays violin plots and statistical values of the LoR distributions, where the SSHM-assisted scenario achieves a mean LoR of 21575 MW·day (std: 8598), compared to 27320 MW·day (std: 10829) in the inspection-based assessment case, indicating not only a 5745 MW·day reduction (21.0\%) in expected loss (VoI) but also a 20.6\% decrease in uncertainty under the presumed earthquake event. These results highlight the significant value of integrating reliable SSHM technologies for timely and precise post-earthquake damage perception, which not only improves expected system resilience but also ensures more predictable recovery trajectories under varying earthquake-induced damage patterns.

\subsection{Sensitivity analysis on VoI}
While the preceding analyses clearly demonstrate the potential benefits of SSHM in improving post-earthquake recovery, they are based on an idealized assumption of fully implemented SSHM on all the EPN components with 90\% detection accuracy and immediate reporting, representing an optimistic best-case for damage perception that is unlikely to be achievable in practice. To offer practical guidance for the design and deployment of SSHM systems in EPNs, this subsection further investigates how specific SSHM performance parameters, such as SSHM assessment accuracy levels, network coverage rates, and reporting delays, affect the effectiveness of SSHM in enhancing post-earthquake resilience.

To this end, a series of sensitivity analyses were conducted by systematically varying the parameter ($a$) controlling the sensing accuracy and the proportion ($p$) of components equipped with SSHM. Specifically, SSHM accuracy levels were tested at $a$ = 0.75, 0.85 and 0.95 representing low, moderate, and high sensing fidelity, respectively. SSHM coverage ratios were varied from 0\% (inspection-based assessment only) to 100\% (full SSHM deployment), with intermediate scenarios ($p$ = 10\%, 30\%, 50\%, 70\%) capturing partial monitoring schemes. Based on the SSHM coverage level, the time required for manually inspecting the remaining unmonitored components was set to 1.8, 1.4, 1.0, and 0.6 days, respectively. In each simulation, SSHM-equipped components were randomly selected according to the specified coverage ratio, and a total of 10000 MC simulations were performed per parameter level, enabling robust statistical assessment of how variations affect the VoI measured by the reduction in expected LoR. 

\begin{figure}
    \centering
    \includegraphics[width=0.75\linewidth]
    { 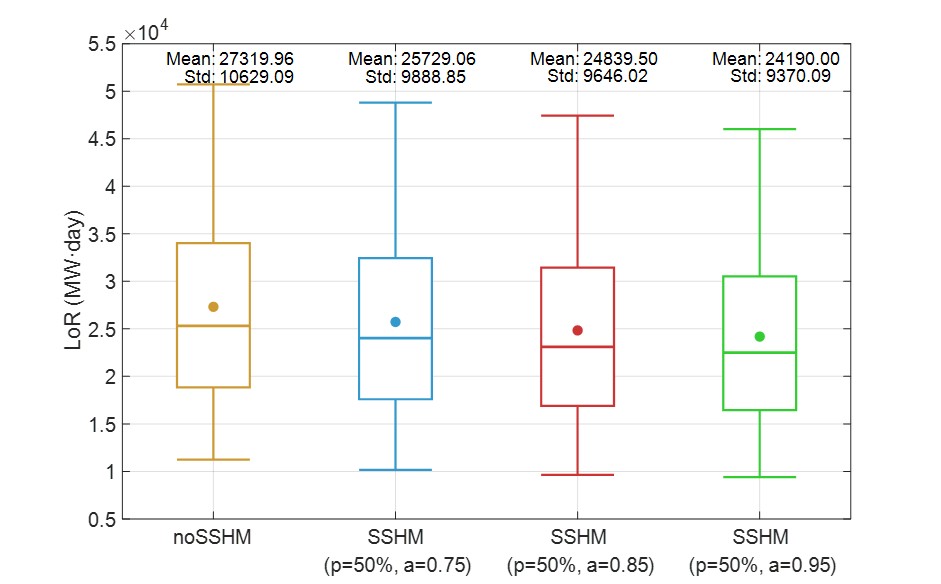}
    \caption{Violin plots of LoR under different SSHM detection accuracies at a fixed coverage ratio $p$=50\%, illustrating the progressive reduction in expected loss and uncertainty as accuracy increases from 0.75 to 0.95. Even at the lowest tested accuracy, SSHM scenarios show substantially lower mean LoR and standard deviation compared to the noSSHM case.}
    \label{detection accuracies}
\end{figure}

Figure \ref{detection accuracies} illustrates the impact of SSHM detection accuracy on LoR distributions under a fixed coverage of $p$=50\%. As SSHM accuracy is increased from 0.75 to 0.95, the mean LoR is observed to decline progressively from 25729 MW·day to 24190 MW·day, accompanied by a reduction in standard deviation from 9889 to 9370 MW·day. These results indicate that higher SSHM accuracy, even at moderate coverage levels, leads to improved expected resilience and reduced uncertainty by minimizing the misclassification of damaged and undamaged components. Moreover, when compared to the noSSHM case, it is observed that even the lowest tested SSHM assessment accuracy ($a$=0.75) achieves a substantial reduction of 1591 MW·day in expected loss, highlighting the significant resilience benefits provided by SSHM regardless of detection accuracy.

\begin{figure}
    \centering
    \includegraphics[width=0.95\linewidth]
    { 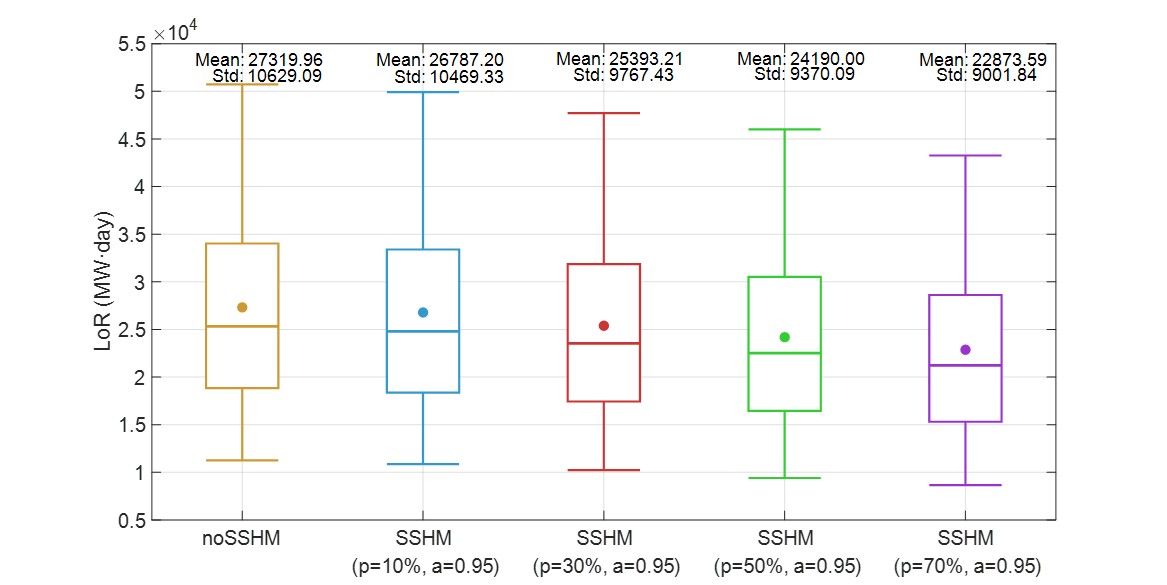}
    \caption{Violin plots of LoR distributions under varying SSHM spatial coverage with fixed detection accuracy ($a$=0.95). As coverage increases from 10\% to 70\%, a substantial reduction in both mean LoR and standard deviation is observed.}
    \label{spatial coverages}
\end{figure}

Figure \ref{spatial coverages} presents the influence of SSHM spatial coverage on LoR distributions with a constant high detection accuracy of a=0.95. As the coverage ratio p is expanded from 10\% to 70\%, the mean LoR declines substantially from 26787 MW·day to 22874 MW·day, and the associated standard deviation correspondingly narrows from 10469 to 9002 MW·day. These findings demonstrate that broader SSHM deployment improves system-wide damage awareness, facilitating more reliable and efficient recovery planning following seismic events. Furthermore, the corresponding VoI for each monitoring setup can also be calculated by comparing the expected reduction in LoR relative to the noSSHM case through Equation (\ref{eq12}). The marginal VoI increase per 10\% in coverage is approximately 697, 602, and 658 MW·day for the p=10–30\%, 30–50\%, and 50–70\% intervals, respectively, indicating that while VoI consistently grows with increased coverage, the marginal benefits show slight variation and an overall trend of diminishing returns, underscoring the importance of balancing sensor deployment density and investment efficiency.

\begin{figure}
    \centering
    \includegraphics[width=0.7\linewidth]
    { 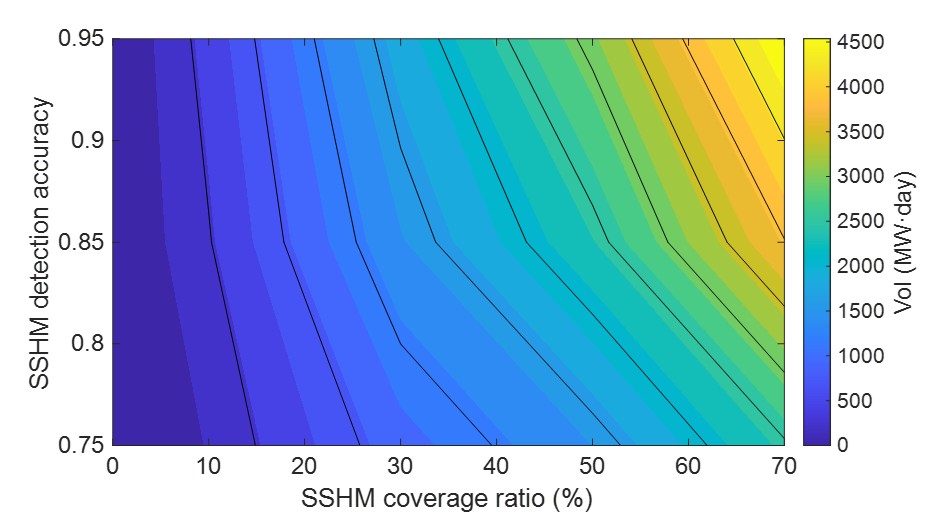}
    \caption{Heatmap of VoI under varying SSHM spatial coverage ($p$) and detection accuracy ($a$) configurations. Contours highlight regions of similar VoI, illustrating how simultaneous increases in coverage and accuracy synergistically enhance expected system resilience.}
    \label{Heatmap of VoI}
\end{figure}

To further investigate the combined effects of SSHM detection accuracy and spatial coverage on system resilience, VoIs were calculated across multiple accuracy–coverage combinations. The results are presented in Figure \ref{Heatmap of VoI}, where intermediate VoI values between tested SSHM accuracy and coverage levels were interpolated to visualize the trends over the practical parameter space. As shown, VoI increases consistently with both higher accuracy and broader coverage, following a nonlinear but monotonic growth pattern. The steepest gains occur when moving from low to moderate accuracy (e.g., $a$ = 0.75 to 0.85) or from sparse to moderate coverage (e.g., $p$ = 10\% to 50\%), while further improvements yield diminishing returns.  The highest VoI is achieved when high accuracy is paired with extensive coverage. Conversely, scenarios with low accuracy but high coverage, and vice versa, achieve only moderate VoI gains, revealing that both sensing fidelity and deployment density must be considered jointly to maximize resilience benefits. The contour lines highlight that comparable VoI levels can be attained through different accuracy-coverage trade-offs, indicating that moderate accuracy with broader coverage can yield resilience improvements similar to those achieved with high accuracy but limited deployment. These findings offer practical guidance for optimizing SSHM system design by balancing accuracy and coverage according to budget constraints and resilience targets. 

As an illustration to further examine the cost-effectiveness of implementing different monitoring strategies in monetary terms, the expected resilience gain VoI was converted into economic value by introducing the Value of Lost Load (VOLL), which represents the unit loss of unserved power. Based on this, a normalized metric, denoted as the VoI-to-Cost Ratio (VCR), was defined as:
\begin{equation}
    \text{VCR}(p, a) = \frac{\text{VoI}(p, a) \times \text{VOLL}}{\text{Cost}(p, a)}
    \label{eq14}
\end{equation}
where VoI(p,a) represents the expected gain in system resilience (MW·day) under a given SSHM coverage rate $p$ and detection accuracy $a$, Cost($p$,$a$) denotes the total investment required for the corresponding monitoring configuration, and VOLL (in USD/MWh) represents the monetary value of each unit of unserved energy, which varies by region and sector ranging from 1000 to 30000 USD/MWh depending on customer type and grid criticality \cite{lacommare2004understanding,van2007overview}. In this study, a representative value of 10000 USD/MWh (240000 USD/MW·day) was adopted, and a fully deployed SSHM system with medium detection accuracy ($a$ = 0.85) was assumed to cost 10 million USD. Then the cost of each other configuration can be scaled linearly based on spatial coverage ratio $p$ and adjusted by a factor of 0.8 for low accuracy ($a$ = 0.75) and 1.2 for high accuracy ($a$ = 0.95), representing the relative expense of sensor precision and data processing. For example, a 50\% coverage with high-accuracy monitoring sensors, i.e., Cost(50\%,0.95), would cost 10 MUSD×0.5×1.2= 6 MUSD. 

\begin{figure}
    \centering
    \includegraphics[width=0.7\linewidth]
    { 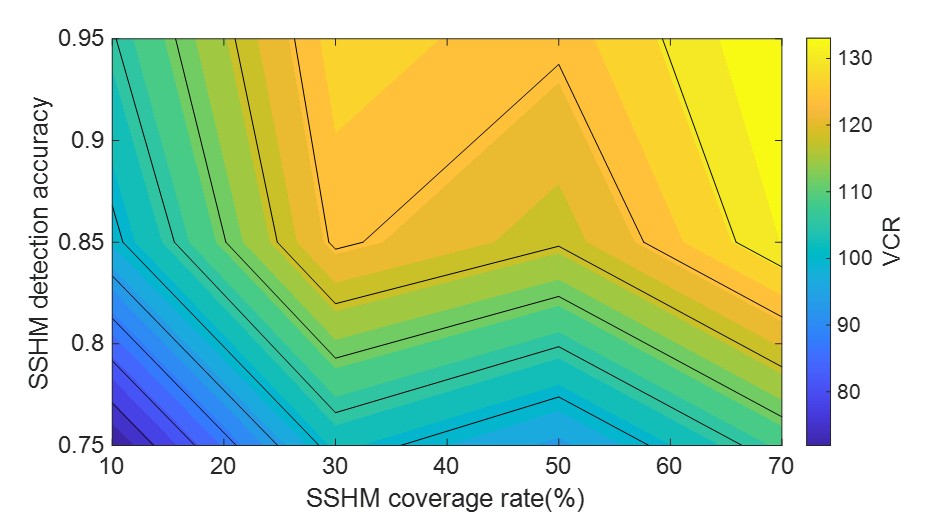}
    \caption{Heatmap of VCR under varying SSHM spatial coverage ($p$) and detection accuracy ($a$) configurations, offering actionable insights for monitoring design under limited budgets.}
    \label{Heatmap of VCR}
\end{figure}

The results are visualized in Figure \ref{Heatmap of VCR}. As shown, low-accuracy configurations consistently perform poorly in cost-effectiveness even under high coverage, as the limited precision undermines the resilience benefits relative to the incurred cost. In contrast, high-accuracy configurations generally yield higher VCR, though the benefits plateau as costs rise disproportionately with increased coverage. The optimal VCR appears in the upper-right corner, where both coverage and accuracy are high. However, this region also corresponds to high total investment levels, which may not be feasible in practice. Notably, the configuration at 30\% coverage with 0.95 detection accuracy emerges as an attractive option, achieving near-optimal VCR values with substantially lower total investment. This highlights a strategically valuable operating point, where targeted sensor deployment with high precision can yield substantial resilience gains per dollar invested, without requiring extensive SSHM system instrumentation. The VCR metric derived from the proposed framework thus provides a practical and scalable approach to guide SSHM design in resilience planning under financial limitations.

\begin{figure}
    \centering
    \includegraphics[width=1\linewidth]
    { 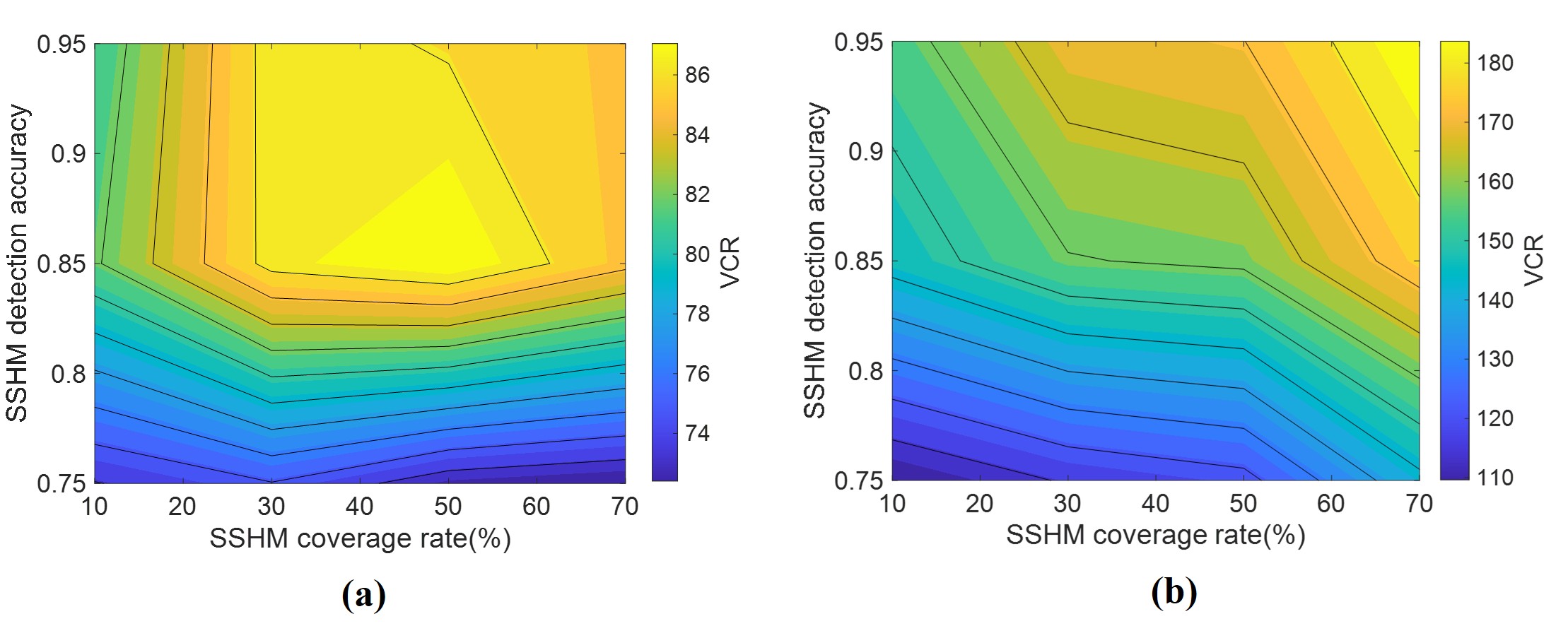}
    \caption{Heatmaps of VCR under varying SSHM coverage and accuracy configurations for earthquake scenarios with magnitudes of (a) 7.5 Ms and (b) 8.5 Ms, respectively, providing robust decision support for optimal monitoring investment.}
    \label{Heatmap of VCRs}
\end{figure}

To evaluate the robustness of SSHM deployment strategies across varying seismic intensities, additional simulations were performed under lower (7.5 Ms), and higher (8.5 Ms) magnitude earthquake scenarios. As illustrated in Figure \ref{Heatmap of VCRs}, VCR values increase consistently with earthquake event severity, reflecting the growing value of monitoring in mitigating more severe disruptions. Notably, the overall spatial distribution of VCR across the SSHM design space remains largely stable, with high detection accuracy ($a$ = 0.95) yielding strong cost-effectiveness regardless of coverage level. In particular, the configuration with $a$ = 0.95 and $p$ = 30\% maintains near-optimal performance under all three seismic scenarios, suggesting that prioritizing sensing accuracy offers a robust and scalable strategy. In contrast, low-accuracy deployments provide limited and less stable returns, particularly under stronger earthquakes. Furthermore, considering that high-magnitude earthquakes typically occur less frequently, monitoring strategies that maintain cost-effectiveness under minor and moderate seismic events—such as the ($a$ = 0.95, $p$ = 30\%) configuration—are particularly valuable. They offer a practical balance between resilience enhancement and financial viability, ensuring preparedness for both frequent moderate shocks and rare but catastrophic events. Future work may incorporate probabilistic seismic hazard models and annualized cost–benefit analysis to further generalize these insights under long-term planning frameworks.

\section{Conclusions}

This study proposes a comprehensive simulation-based framework to quantify the resilience-based Value of SSHM in supporting the post-earthquake recovery of EPNs. The key contributions of this study are summarized as follows:

First, it advances conventional resilience assessment and post-earthquake recovery modeling by explicitly incorporating damage assessment uncertainty through a perception-aware recovery simulation. By representing SSHM performance via confusion matrices, the proposed approach captures the impact of sensing accuracy and misclassification on repair prioritization and resilience outcomes, overcoming the unrealistic assumption of perfect damage knowledge common in existing models.

Second, it develops a comprehensive, modular, scenario-based computational framework that integrates probabilistic damage simulation, SSHM perception modeling, recovery scheduling dynamics, and quantitative resilience evaluation. This end-to-end framework enables systematic, resilience-oriented analysis of different SSHM configurations under practical resource and operational constraints, bridging the gap between damage detection research and actionable system-level recovery planning.

Third, it conducts comprehensive sensitivity analyses on practical combinations of SSHM sensing accuracy and coverage, providing quantitative insights into their effects on key resilience metrics LoR. Furthermore, by combining VoI estimates with deployment costs, a normalized cost-effectiveness metric VCR is formulated to examine the cost-effectiveness of different monitoring strategies in monetary terms, thus offering actionable insights for monitoring design under limited budgets.

The proposed framework was demonstrated through a case study on the IEEE 24-bus Reliability Test System. Results show that improved damage awareness enabled by SSHM significantly accelerates recovery and reduces LoR compared to the noSSHM baseline. Sensitivity analysis on SSHM parameters and cost-effectiveness analysis on VCR reveal that optimal monitoring deployment often favors high-sensing accuracy with moderate coverage—offering strong economic returns without necessitating full SSHM system instrumentation. Such analysis can provide quantitative insights into the system-level value of SSHM and support evidence-based investment in sensing technologies for critical infrastructure resilience enhancement with available funding. 

Overall, this study provides a rigorous and scalable foundation for assessing the system-level benefits of SSHM in the post-earthquake context. By emphasizing realism, uncertainty quantification, and actionable insights, this study complements prior work in infrastructure resilience by establishing a quantitative bridge between SHM technologies and resilience-centric infrastructure management, while also opening new avenues for cross-disciplinary research at the intersection of sensing, optimization, and disaster recovery. Future research will focus on damage confusion matrix calibration, multi-hazard extensions, real-time data assimilation, and decision-making under deep uncertainty. 

\section*{Acknowledgments}
The research was conducted at the Singapore-ETH Centre, which was established collaboratively between ETH Zurich and the National Research Foundation Singapore, and CNRS@CREATE through the DESCARTES program, both research supported by the National Research Foundation, Prime Minister’s Office, Singapore under its Campus for Research Excellence and Technological Enterprise (CREATE) programme. Prof. Chinesta (F.C) also acknowledges the support of the Chimera RTE research chairs at Arts et Metiers Institute of Technology (ENSAM). Prof. Chatzi gratefully acknowledges the funding from Horizon Europe under the program HORIZON-CL5-2023-D4-02-01 for project INBLANC - INdustrialisation of Building Lifecycle data Accumulation, Numeracy and Capitalisation, Grant Number: 101147225.

\bibliographystyle{unsrt}  
\bibliography{refs}

\end{document}